# TITLE

Automated assessment of disease severity of COVID-19 using artificial intelligence with synthetic chest CT

# AUTHORS

Mengqiu Liu[1,#], Ying Liu[1,#], Yidong Yang[2], Aiping Liu[3], Shana Li[1], Changbing Qu[1], Xiaohui Qiu[4], Yang Li[5], Weifu Lv[1], Peng Zhang[6], Jie Wen[1,*]

1. Department of Radiology, The First Affiliated Hospital of USTC, Hefei, China, 230001
2. School of Physical Sciences, University of Science and Technology of China, Hefei, China, 230001
3. Electronic Science and Technology, University of Science and Technology, Hefei, China, 230001
4. Radiology, Bozhou people's hospital, Bozhou, China, 236800
5. Radiology, The first affiliated hospital of Bengbu medical college, Bengbu, China, 233000
6. AItour, Japan

#The two authors contributed equally.

*Corresponding author:

Jie Wen

Radiology, The first affiliated hospital of USTC, Hefei, China, 230001

Tel.       +86-0551-62283486

Mail:     jiewen@ustc.edu.cn

## ABSTRACT

**Background**

Triage of patients is important to control the pandemic of coronavirus disease 2019 (COVID-19), especially during the peak of the pandemic when clinical resources become extremely limited.

**Purpose**

To develop a method that automatically segments and quantifies lung and pneumonia lesions with synthetic chest CT and assess disease severity in COVID-19 patients.

**Materials and Methods**

In this study, we incorporated data augmentation to generate synthetic chest CT images using public available datasets (285 datasets from "Lung Nodule Analysis 2016"). The synthetic images and masks were used to train a 2D U-net neural network and tested on 203 COVID-19 datasets to generate lung and lesion segmentations. Disease severity scores (DL: damage load; DS: damage score) were calculated based on the segmentations. Correlations between DL/DS and clinical lab tests were evaluated using Pearson's method. A p-value < 0.05 was considered as statistical significant.

**Results**

Automatic lung and lesion segmentations were compared with manual annotations. For lung segmentation, the median values of dice similarity coefficient, Jaccard index and average surface distance, were 98.56%, 97.15% and 0.49 mm, respectively. The same metrics for lesion segmentation were 76.95%, 62.54% and 2.36 mm, respectively. Significant ($p \ll 0.05$) correlations were found between DL/DS and percentage lymphocytes tests, with r-values of -0.561 and -0.501, respectively.

**Conclusion**

An AI system that based on thoracic radiographic and data augmentation was proposed to segment lung and lesions in COVID-19 patients. Correlations between imaging findings and clinical lab tests suggested the value of this system as a potential tool to assess disease severity of COVID-19.

**Summary**

Data augmentation was incorporated in an AI system that was used to segment lung and pneumonia lesions in COVID-19 patients. Quantitative analysis of the imaging data provides a potential tool for disease severity assessment of COVID-19 patients.

**Key Results**

- An AI system was proposed to segment lung and lesions in 203 COVID-19 patients. By incorporating data augmentation, synthetic chest CT images were generated using public available datasets to avoid labor-intensive annotations.
- Automatic lung and lesion segmentations were compared with manual annotations. For lung segmentation, the median values of dice similarity coefficient, Jaccard index and average surface distance, were 98.56%, 97.15% and 0.49 mm, respectively. The same metrics for lesion segmentation were 76.95%, 62.54% and 2.36 mm, respectively.
- Significant ($p \ll 0.05$) correlations were found between DL ($r = -0.561$) or DS ($r = -0.501$) and percentage lymphocytes tests.

## INTRODUCTION

Up to September 25, 2020, 32,110,656 coronavirus disease 2019 (COVID-19) cases have been reported worldwide, and 980,031 cases were confirmed death, resulting a fatal rate up to 3.05% (1). Although over 169 COVID-19 vaccine candidates are under development, with 26 of them in the human trial phase, efficient diagnosis and assessment are crucial to tracking the disease and mitigating its spread.

Currently, Reverse Transcription-Polymerase Chain Reaction (RT-PCR) test is considered as the gold standard of COVID-19 diagnosis. (2) However, the sensitivity of RT-PCR assay was reported to be as low as 60-89%, (3-5) which might lead to false negative diagnosis, especially in the early phase of infection. On the other hand, imaging techniques, such as chest X-ray and thoracic CT, have been demonstrated to have improved sensitivity compared to RT-PCR, and can provide huge assistance to clinicians. A recently study (6) conducted in France showed that, chest CT sensitivity and specificity for COVID-19 diagnosis can reach up to 90%, and, as high as 90% of CT-positive and RT-PCR-negative COVID-19 patients had later been confirmed by positive repeat RT-PCR results.

Severity assessment is crucial for patient triage and treatment planning, and can help to optimize the allocation of limited clinical resources, particularly during the peak of the pandemic. (7-13) Lymphopenia is a potential biomarker for disease severity, which is known to be prevalent among patients with COVID-19. Previous studies have reported that peripheral blood lymphocyte cell count decline is predictive of disease severity and death. (7-12, 14) It was also shown that COVID-19 patients with lymphopenia could benefit from treatment with recombinant human granulocyte colony-stimulating factor (rhG-CSF). (15)

Since the outbreak of the pandemic, artificial intelligence (AI) techniques have been used actively in fighting COVID-19, such as disease diagnosis, severity assessment and so on. Lung and lesion segmentation is usually an essential step in disease severity assessment of COVID-19. Currently, many popular networks – such as U-Net (16), UNet++ (17) and VB-Net (18) – have been used for lung and lesion segmentation. However, considerable large datasets with manual annotation are required to train a robust segmentation network, which is labor-intensive and time-consuming. Previously, several studies (18, 19) incorporated human knowledge into the training process. This human-in-the-loop strategy, together with other weakly-supervised machine learning methods (20, 21), are highly demanded when annotated medical images are lacking in COVID-19 studies.

Here we propose an AI system that can efficiently segment lung regions and COVID-19 pneumonia lesions, namely ground glass opacities and consolidations. By using data augmentation techniques, we were able to create "pseudo" lesions as labels in the lung regions on public available CT datasets. This minimizes the need for large number of COVID-19 data and labor intensive manual annotations. It also allows us to quantify disease severities in COVID-19 patients. The disease severity scores calculated from CT images showed significant correlations with percentage lymphocytes (LYM%), indicating the reliability of these scores as useful measures for disease severity assessment.

## MATERIALS AND METHODS

### Datasets

As a retrospective study, it was approved by the local ethics committees to waive the informed consent of the patients. A total number of 203 lung CT datasets were collected from multiple hospitals in Anhui province, China. All images were collected from COVID-19 patients with positive RT-PCR test. CT scans were collected on scanners from different manufacturers. Lab tests, including white blood cell count (WBC) and percentage lymphocytes (LYM%), were also collected for each patient. Detailed information of demographic, CT scan parameters and lab test results are shown in Table 1.

Table 1 Summary of demographic, CT scan parameters and lab tests results

| Demographic | | | | |
|---|---|---|---|---|
| Total number | 203 | | | |
| Sex (female/male) | 95/108 | | | |
| Age (mean±SD) | 42.65±17.31 | | | |
| CT scans | | | | |
| Manufacturer | SIEMENS | NEUSOFT | GE | TOSHIBA |
| Number of scans | 107 | 77 | 18 | 1 |
| Slice thickness (mm) | ≤0.8 | 0.8-2 | 2-5 | > 5 |
| Number of slices | 129 | 15 | 54 | 5 |
| Lab test | | | | |
| WBC (mean±SD, ×10^9/L) | 5.35±2.10 | | | |
| LYM (mean±SD, %) | 27.84±13.13 | | | |

Public available lung CT datasets (LUNA16 – Lung Nodule Analysis 2016) were also used in this study. To minimize the influence of pulmonary nodules on our analysis, we removed all the datasets that had pulmonary nodules > 3mm, which retained a total number of 285 datasets collected on scanners with different manufacturers and slice thicknesses. These images were then used to generate synthetic chest CT with pseudo lesions inside the lung regions and mimicked the abnormalities in COVID-19 cases.

**Lung & lesion segmentation**

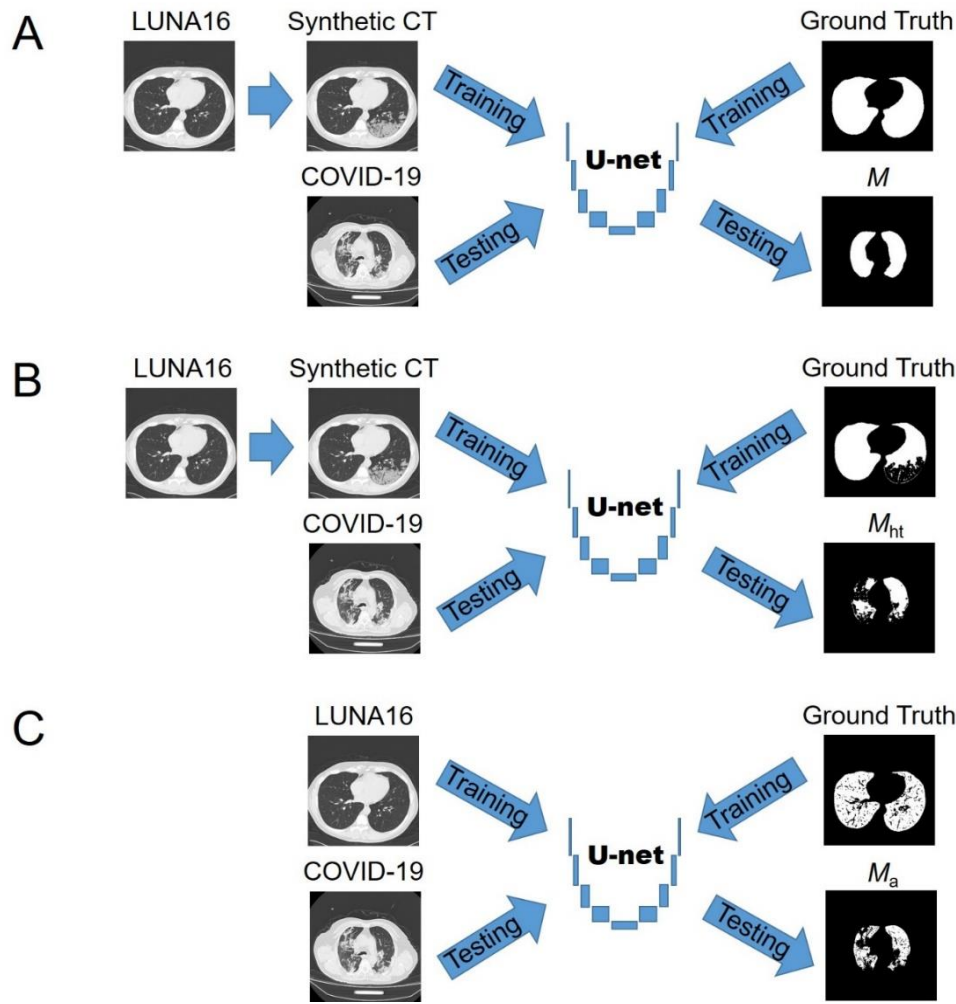

Figure 1 Overview of the AI system used for lung and lesion segmentation. A. "Normal" CT images from LUNA16 dataset were used to generate synthetic CT with "pseudo" lesions. Synthetic CT images and original LUNA16 labels were used to train a standard 2D U-net. Output model was then applied on COVID-19 images to generate "lung" masks – $M$. B. "Healthy" lung tissue masks were also generated when creating the synthetic CT images, and used as ground truth to train a standard 2D U-net. Output model was then applied on COVID-19 images to produce "healthy" lung tissue masks – $M_{ht}$. C. "Pulmonary air" masks – $M_a$ – were also generated based on magnitude thresholding, as described in the main text. Original LUNA16 images and generated $M_a$ were used to train a standard 2D U-net, which was then tested on COVID-19 images. All three masks were combined together to create fine segmentations of lungs and lesions, which were later used for disease severity assessment.

An Overview of the AI system is shown in Figure 1. "Normal" (with nodule diameters < 3mm) CT images (n=285) from LUNA16 dataset were used to generate synthetic CT images with "pseudo" pneumonia lesions. First of all, a "seed" was placed on the boundary between lung and tissue, and used as a starting point. Secondly, a "random walk" strategy was used to spread the "lesion" area. "Lesion" voxels were randomly assigned CT values between -650 HU and -180 HU to mimic COVID-19 pneumonia lesions. The number of walking steps were limited not to exceed the size of the lung masks. A mask of "healthy" lung tissue – $M_{ht}$ – was also generated for each dataset. The generated images and masks (as shown in Figure 2) were then used for U-net training.

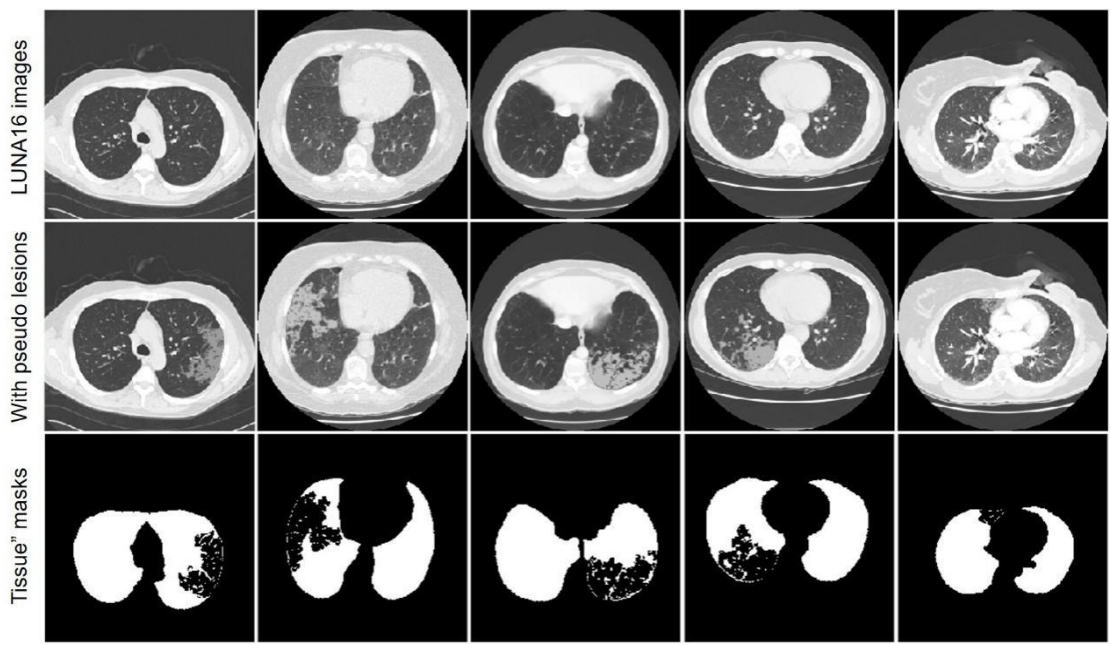

Figure 2. Examples of LUNA16 images (top row) and generated synthetic images with pseudo lesions (middle row). Masks of "healthy lung tissues" were shown on the bottom. Images were collected from 5 LUNA16 datasets.

"Normal" LUNA16 images were also used to generate "pulmonary air" masks – $M_a$. For each 3D CT dataset, a histogram of CT values in the whole lung regions was plotted (Figure 3). The upper half of the highest peak of the distribution was fitted to a normal distribution. The fitting parameters μ (mean) and σ (standard deviation) were used to calculate a threshold = μ+2.58σ. A binary "pulmonary air" mask was created by setting all lung regions that had CT values smaller than the threshold to 1.

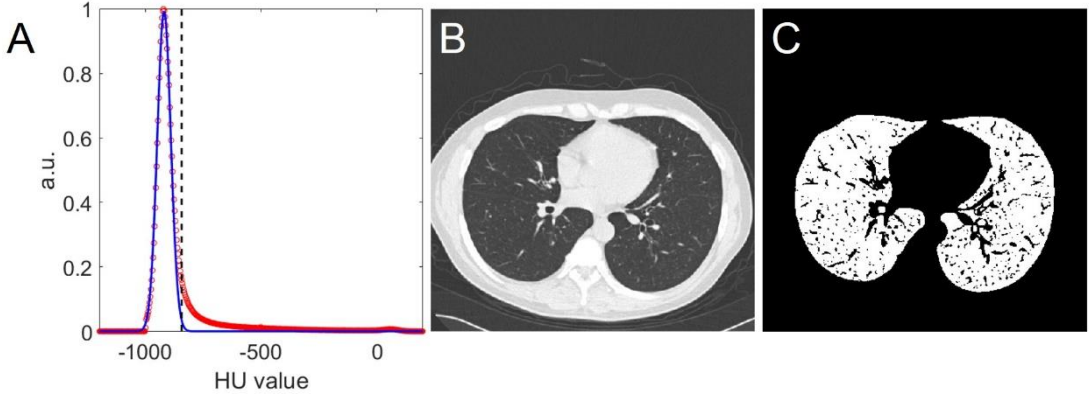

Figure 3 Generation of "pulmonary air" masks – $M_a$. Histogram (A) of CT HU values in lung regions was calculated from "normal" LUNA16 chest CT images (B). Red circles represent real distribution. A normal distribution (blue line) is used to fit the upper half of the highest peak of the distribution. The black dashed line represents a threshold which equals μ+2.58σ, where μ and σ are the mean and standard deviation of the fitted normal distribution. The "pulmonary air" mask (C) only kept regions that have CT values smaller than the threshold.

All the "LUNA16" images and masks were put into a standard 2D U-net for training. The U-net was adapted from https://github.com/zhixuhao/unet and was implemented with Keras functional API. Three models were trained: model 1 that provided masks of the whole lung regions – $M$; model 2 that provided rough masks of "healthy" lung tissue – $M_{ht}$; and model 3 that provided masks of the "pulmonary air" regions – $M_a$. By combining these 3 masks, a fine segmentation of the lung and lesion regions – $M_s$ – was calculated:

$$M_s = M \cap (1 - M_{ht}) \cap (1 - M_a) \qquad (1)$$

where $M$, $M_{ht}$ and $M_a$ represent masks of the whole lung, "healthy" lung tissue and "pulmonary air", respectively. The segmentations were then used in disease severity assessment.

**Manual annotation**

Algorithm-generated segmentations were compared with manual annotations. To reduce the heavy annotation burden, we employed a human-in-the-loop strategy, similar to what other researchers did in (18, 22). Three individuals performed refinements to algorithm-generated results to produce manual annotations. Dice similarity coefficients (DSC), Jaccard index (JI) and average surface distance (ASD) were calculated to evaluate the performance of the automated segmentations. The evaluation metrics were defined as follows:

$$DSC(A,B) = \frac{2|A \cap B|}{|A| + |B|} \quad (2)$$

$$JI(A,B) = \frac{|A \cap B|}{|A \cup B|} \quad (3)$$

$$ASD(A,B) = \frac{1}{2} \left( \frac{\sum_{a \in S(A)} \min_{b \in S(B)} \|a - b\|}{|S(A)|} + \frac{\sum_{b \in S(B)} \min_{a \in S(A)} \|b - a\|}{|S(B)|} \right) \quad (4)$$

where A and B represent regions of automated and manual segmentations, respectively. S(A) and S(B) represent the set of surface points of A and B, respectively.

**Severity assessment**

Two abnormality measures were developed:

$$DL = \frac{V_L}{V}$$
$$DS = MCT_L \quad (5)$$

where $DL$ and $DS$ represent damage load and damage score, respectively. $V_L$ is the volume of the lesions and $V$ represents volume of the entire lung. $MCT_L$ represents median CT HU values calculated from the lesion regions.

**Statistical analysis**

All statistical analyses were performed using MATLAB (The MathWorks, Inc).

Pearson's correlation was used to evaluate the correlation between DL or DS and clinical measures (WBC and LYM%). Correlation coefficient (r) and p-values (p) were calculated. In all cases, a p-value < 0.05 was considered as statistically significant.

## RESULTS

### Lung segmentation

Results of lung segmentation are shown in Figure 4. Red contours indicate segmentations obtained using a U-net trained on the original LUNA16 images, which introduced significant segmentation errors. On the contrary, blue contours illustrated segmentations obtained using a U-net trained on the synthetic images with generated "pseudo" lesions. Adding this data augmentation step greatly reduced errors and improved the performance compared with manual annotations. Achieved median values of DSC, JI and ASD were 98.56%, 97.15% and 0.49 mm, respectively.

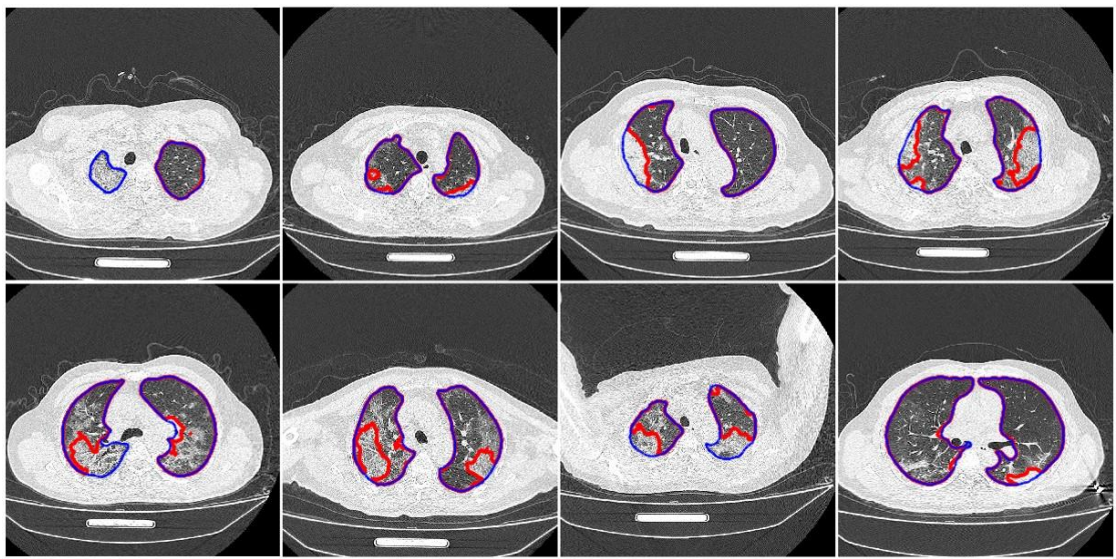

Figure 4 Illustration of the performance of lung segmentation method that used a U-net trained on original LUNA16 images without (red contours) and with "pseudo" lesions (blue contours). Images were collected from 8 different COVID-19 patients.

### Abnormality segmentation

Lesion segmentations are illustrated in Figure 5, as marked by the red contours.

Complex appearances of pneumonia lesions often make it difficult to obtain precise segmentations. Incorporating data augmentation technique greatly improved the segmentation results and reduced the requirement of labor-intensive manual annotations. For all 203 COVID-19 patients, the median values of DSC, JI and ASD were 76.95%, 62.54% and 2.36 mm, respectively. These results are comparable with previous studies that used manual annotations for training. (22-24)

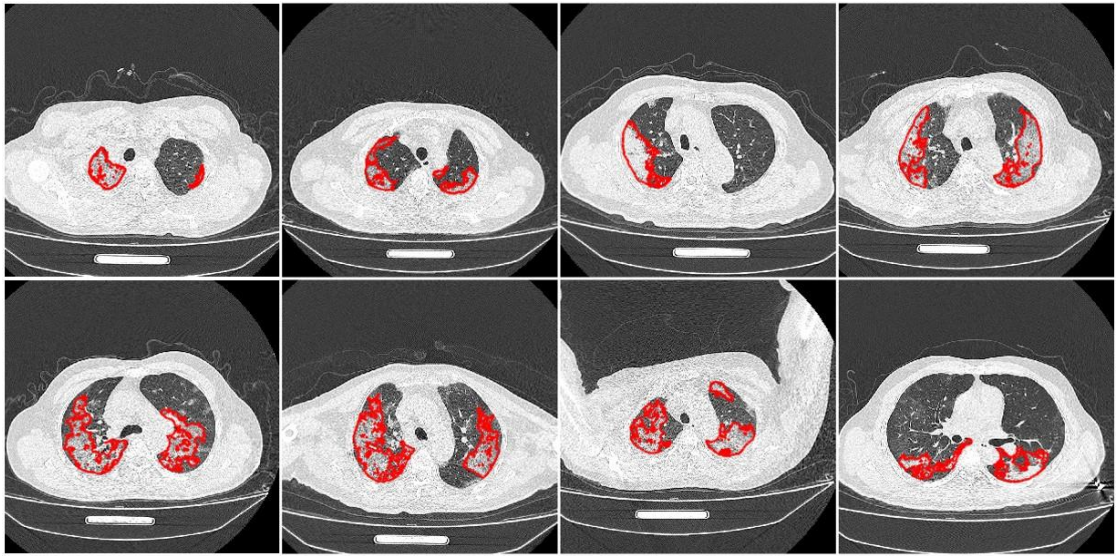

Figure 5 Illustration of lesion segmentations. Red contours indicate the lesion regions. The same images were used here as in Figure 4.

**Correlation between disease severity scores and clinical measures**

A WBC count can detect hidden infections within the body and Lymphopenia is known to be prevalent among patients with COVID-19. Previous studies have reported that peripheral blood lymphocyte cell count decline is predictive of disease severity and death. (7-12, 14) Our results showed that DL and DS scores that were evaluated from thoracic CT scans correlated with LYM% measures, indicating that DL and DS can be used as potential measures for disease severity assessment in COVID-19 patients.

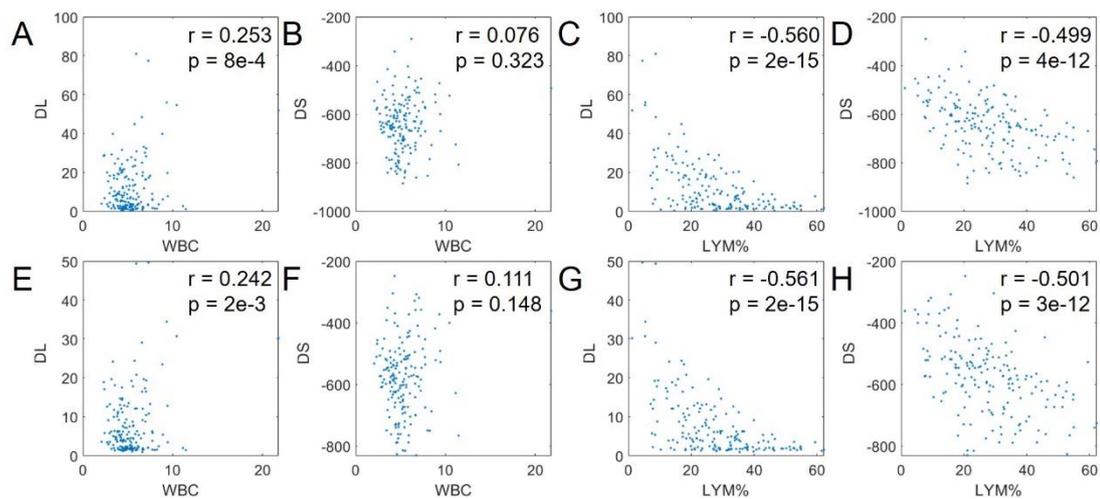

Figure 6 Correlations between DL or DS and clinical tests (WBC and LYM%). Manual (top row: A – D) and automatic (bottom row: E – H) annotations were used to calculate DL and DS. The r-values and p-values of the correlations are listed in each panel. Each point represents an individual subject.

**DISCUSSION**

In this study, we proposed an AI system that can provide accurate segmentations of lung and COVID-19 related pneumonia lesions which appear as ground glass opacities and consolidations. By using data augmentation techniques, we were able to create "pseudo" lesions in the lung regions on public available CT datasets. This strategy minimizes the need for large number of COVID-19 training samples and labor intensive manual annotation. The auto-segmentations can be directly used to quantify disease severities in COVID-19 patients.

AI techniques provide an efficient and accurate way for patient triage and disease diagnosis, and can help to optimize the allocation of medical resources. Fast and precise assessment of disease severity and progression allows physicians to quickly determine treatment strategies and is the most important step in ensuring the quality of follow-up treatment. Automatic segmentation of lung and lesion regions is an important initial step for rapid assessment of pulmonary infections and disease progression in COVID-19 patients. Although AI has been used to generate lung and lesion segmentations in patients with COVID-19, most of the previous applications heavily rely on training data that require labor-intensive manual annotations. The manual contouring process is very time consuming and there exists considerable inter-physician variations heavily depending on personal experience. Therefore, it is almost impossible to obtain a unified and objective evaluation.

The contribution of this study is that we introduced data augmentation techniques to generate "pseudo" lesions to mimic COVID-19 abnormalities on the CT images using public available dataset – LUNA16. This greatly reduced the need of manual annotations and allowed the neural network to be trained on existing public datasets. The results showed that the proposed automatic system used in this study can achieve satisfying segmentations of lung and lesion, which is comparable with existing methods that requrie manual annotations in their training data. The correlations between disease severity scores and clinical lab tests further indicate that this system can provide measures for disease severity assessment.

There is still room for improvement in the proposed system in the future. In current study, we used a "random walk" strategy to generate "pseudo" lesions. Although comparison with manual annotations has validated this strategy, more advance technique, such as semi-supervised learning, (25, 26) can be used to further minimize the distribution discrepancy between synthetic and real datasets.

In conclusion, we proposed an AI system that can efficiently segment lung regions and COVID-19 related pneumonia lesions. By using data augmentation in the training procedure, we were able to create synthetic chest CT images with "pseudo" lesions to mimic COVID-19 abnormalities and avoid the requirements of labor-intensive manual annotations. The disease severity scores calculated based on auto-segmentations were used to quantify disease severity and showed significant correlations with percentage lymphocytes (LYM%) tests, indicating that this system can provide potential measures for disease severity assessment in COVID-19.

**ACKNOWLEDGMENTS**